\documentclass{ws-ijmpa}
\usepackage[super,compress]{cite}
\usepackage{graphicx}
\usepackage{mathabx}
\begin{document}

%
\catchline{}{}{}{}{}
%

\title{PLAYING WITH  THE  ENVELOPING  ALGEBRA  OF SUPERSYMMETRY
}

\author{E. Cattaruzza}

\address{INFN, Sezione di Trieste, Trieste, Italy\\
Enrico.Cattaruzza@ts.infn.it}

\author{E. Gozzi}

\address{Theoretical Section Dept. of Physics, University of Trieste, Trieste, Italy\\
INFN, Sezione di Trieste, Trieste, Italy\\
gozzi@ts.infn.it}

\maketitle

\begin{abstract}
In this paper we show how to obtain from a scalar superfield its first component via a similarity transformation.
We prove that  in $D=4$ the generators of this similarity transformation live in the enveloping algebra of supersymmetry
while for  $D=1$ they belong to the basic algebra.

\keywords{Supersymmetry; Superfield.}
\end{abstract}

\ccode{PACS numbers: 11.30.Pb, 31.15.xK}


\section{INTRODUCTION}

\par Supersymmetry \cite{ref1.1,ref1.2,ref1.3} is, in our opinion, one of the most interesting idea which has appeared in physics over the last  50 years. It is the only other space-time symmetry, beside the Poincar\'e  invariance, which has been proposed for  flat space-time. Moreover it has a unique role in taming the ultraviolet divergences of quantum field theories \cite{ref2.1,ref2.2}. 

\par Unfortunately so far there is no experimental indication of it. It may be that the phenomenological models that have been tested are not the ones which nature uses. But it may also be that there are further theoretical aspects of supersymmetry  (SUSY) which have not been investigated in enough details. This paper is a modest attempt in this direction. 
\par
SUSY is a symmetry which put in the same multiplet fields of different statistics, bosons and fermions. A short manner to write up all these fields in a single multiplet is via the concept \cite{ref3} of superfield. This concept was crucial also in order to bring to light the geometry hiding behind SUSY. This geometry is  related to the concept of superspace which is an extension via grassmannian coordinates of ordinary space-time. The superfields usually have a first component which is a known physical field like the electron field or the quark or the Higgs, while the other components are partner fields which should somehow appear in nature. 
\par The manner to obtain the first component from the superfield is to bring to zero the grassmannian partners of space-time. This is a very rude manner to proceed which does not throw any light on the interplay between the superfield and its first component which is usually also the only physical component discovered so far. In this paper we shall study another manner to go from the superfield to its first component. It is based on a similarity transformation on the superfield. The generators of this transformation belong to the SUSY algebra in the case of 1-dimensional system and to the enveloping algebra in the 4-dim case.    
\par For the 1-dim case the similarity transformation is basically a sort of change of picture from the Heisenberg to the Schr\"{o}dinger one with respect to the grassmannian partners of time. In the 4-dim case we have not succeeded in giving such a clear-cut interpretation. 
\par The paper is organised as follows: in Sect.\ref{Sect2} we analyse a particular 1-dim. supersymmetric model  which is basically a modern way to rewrite classical mechanics. In Sect.\ref{Sect3} we analyse the 4-dim Wess-Zumino\cite{ref1.1} model and its  associated\cite{ref3} scalar superfield. In this case first we perform, like for the 1-dim case, a similarity transformation which is a field theory implementation of translation in the grassmannian part of superspace. We prove that this transformation does not bring the superfield to its first component but to the set of its scalar components. In Sect.\ref{Sect4} we apply to this first transformation a second one but built with elements  of the enveloping algebra of SUSY. This set of combined transformations brings the superfield to its first component.
\par
Some long and detailed calculations are confined to few appendices.   
\section{1-DIMENSIONAL CASE }\label{Sect2} 
The 1-dim model we will analyse in this section made its appearance in a path-integral approach to classical mechanics reviewed in ref.\citen{ref4}. Its lagrangian is the following one:    
\begin{equation}
\widetilde{\mathcal{L}}=\lambda_a\,\dot\varphi^a+i\,\bar c_a\,\dot{c}^a-\lambda_a \omega^{ab} \frac{\partial H}{\partial \varphi^b}- i \bar{c}_a \omega^{ac} \frac{\partial^2 H}{\partial \varphi^c \partial \varphi^b} c^b.
\label{eq:6-1}
\end{equation}
The "fields" $\varphi^a(t)$ are the $2n$-phase space coordinates of a classical point particle
\[\varphi^a = \left( q^1,\cdots , q^n,p^1,\cdots ,p^n \right) ,\,a=1,\cdots,2n.
\]
while the $\lambda_a(t),\,c^a(t),\,\bar c_a(t)$ are auxiliary coordinates  depending  on  the 1-dim. time variable $t$. We point out that the variables $c^a,\bar c_a$ have odd grassmannian parity\cite{ref7} . Also on these auxiliary variables  the index "$a$"  runs from 1 to $2\,n$ . The symbol $\omega^{ab}$ indicates  the symplectic matrix of the original $\varphi^a$-space:
\begin{equation*}
	\omega^{ab}=\left(\begin{matrix}
 0 & \mathbb I\\
 -\mathbb I & 0 
\end{matrix}
\right)
\end{equation*}
and $H(\varphi)$ is a function of $\varphi$ which plays  the role of a superpotential in supersymmetry but originally was the Hamiltonian of the classical system living in the $\varphi^a$-space. The readers interested in the relation between $\tilde{\mathcal L}$ and the original classical system moving in the $\varphi$-space can consult ref.\citen{ref4}.
Here let us only mention that the $c^a$ turn out to be the basis of the differential forms on $\varphi$
-space, $\bar c_a$ the basis of the totally anti-symmetric tensors and that $\tilde{\mathcal L}$ is related to the Lie-derivative of the Hamiltonian flow\cite{ref6} .
As $\tilde{\mathcal L}$ makes its appearance in a path-integral so we can calculate \cite{ref5.1,ref5.2} the graded commutators \cite{ref7} among the $8\,n$ variables $(\varphi^a,\,\lambda_a,\,c^a,\,\bar c_a)$. The only non-zero graded commutators are the following ones:
\begin{equation}
	\begin{matrix}
\left[\varphi^a,\lambda_b\right] & =& i\,\delta_b^a\\
\left[c^a,\bar c_b\right] & =& \delta_b^a.\\
\end{matrix}
\label{eq:8-1}
\end{equation}
The Hamiltonian associated to the Lagrangian (\ref{eq:6-1}) is: 
\begin{equation}
	\mathcal{\tilde H}=\lambda_a \omega^{ab} \frac{\partial H}{\partial \varphi^b}+ i \bar{c}_a \omega^{ac} \frac{\partial^2 H}{\partial \varphi^c \partial \varphi^b} c^b
	\label{eq:8-2}
\end{equation}
and it is easy to prove that it commutes with the following charges:
\begin{equation}
	\begin{matrix}
Q&\equiv & i\,c^a\,\lambda_a\\
\widebar Q&\equiv&i\,\bar c_a\,\omega^{ab}\,\lambda_b\\
Q_H&=&Q-c^a\,\partial_aH\\
\widebar Q_H&=&\widebar Q+\bar c_a\,\omega^{ab}\,\partial_bH.
\end{matrix}
\label{eq:8-3}
\end{equation}
Besides commuting with $\tilde{\mathcal H}$ it is easy to check that 
\begin{equation}
  \left[Q_H,\widebar Q_H\right]=2\,i\,\widetilde{\mathcal H}.
  \label{eq:8-4}
\end{equation}
So the $Q_H,\,\widebar Q_H$ are generators of an $N=2$ supersymmetry\cite{ref8} . There are other charges which commute \cite{ref8} with $\widetilde{\mathcal H}$ but we are not interested in them in this paper. 
\par
An important tool introduced in ref.\citen{ref3}  for supersymmetry is the one of superfield. This is basically a manner to put in the same multiplet all the $8\,n$ variables $(\varphi^a,\,\lambda_a,\,c^a,\,\bar c_a)$. First we extend the space $t$ via two grassmannian partners $(\theta,\bar \theta)$ to what is called a superspace $(t,\,\theta,\,\bar \theta)$. Next we introduce a function $\Phi^a(t,\theta,\bar \theta)$ in this superspace defined as follows: 
\begin{equation}
  \Phi^a(t,\theta,\bar \theta)\equiv \varphi^a(t)+\theta\,c^a(t)+\bar\theta\,\omega^{ab}\,\bar c_b(t)+i\,\theta\,\bar \theta\,\omega^{ab}\,\lambda_b.
\label{eq:9-1}
\end{equation}
In this superspace it is easy to give to the charges a differential-operator representation.
This is done in the following manner. Let us call the super-space $(t,\,\theta,\,\bar \theta)$ as {\it base-space} and $(\varphi,\,\lambda,\,c,\,\bar c)$ as {\it target-space}. It is then natural to think  that the transformations on the {\it target-space} generated by the charges of eq.(\citen{ref2.1}) are induced by transformations on the {\it base-space} generated by some differential operator. The simplest request we can make on the superfield $\Phi$ is that it is a scalar under the transformations generated by the operators in eq.(\citen{ref2.1})  which means : 
\begin{equation}
\Phi'^a(t',\theta',\bar \theta')=\Phi^a(t,\theta,\bar \theta), 
\label{eqinserzione:7}  
\end{equation}
where the $\Phi'$ indicates the superfield built out of the transformed target-space $(\varphi',\,\lambda',\,c',\,\bar c')$.
If we indicate with $\mathcal O$ any of the charges in eq.(\citen{ref2.1}) and with $\widehat{\mathcal O}$ the associated differential operator in the base-space, the relation (\ref{eqinserzione:7}) is satisfied if and only if the following equation holds:
\begin{equation}
\left[\mathcal O,\Phi^a\right]=-\widehat{\mathcal{O}}\,\Phi^a.
\label{eqinserzione:8}  
\end{equation}
The $\mathcal{O}$ on the L.H.S. of (\ref{eqinserzione:8}) acts on the target space while the $\widehat{\mathcal{O}}$ on the R.H.S. acts on the base-space. As we have the explicit  form of the various charges $Q$ , $\widebar Q$, etc. of eq.(\citen{ref2.1}) it is easy to derive from eq. (\ref{eqinserzione:8}) the expressions of the various $\widehat Q$, $\widehat {\widebar Q}$ etc. associated to those charges. They are: 
\begin{equation}
  \begin{matrix}
  \widehat Q&=&-\cfrac{\partial}{\partial \theta}\equiv -\partial_{\theta}\\
 \widehat{\widebar Q} &=&+\cfrac{\partial}{\partial\bar \theta} \equiv\partial_{\bar{\theta}}\quad\\
\,\,  \widehat Q_H&=& -\partial_{\theta}-\bar \theta\,\partial_t\,\,\\
\,\,\widehat {\widebar Q}_H&=& \partial_{\bar \theta}+ \theta\,\partial_t. 
\label{eq:10-2}  
\end{matrix}
\end{equation}
We would like to stress that the $\mathcal O$ and $\widehat{\mathcal{O}}$ have the same algebra but are reppresented in two different spaces: the  $\mathcal O$ in the {\it target-space} and the $\widehat{\mathcal{O}}$ in the {\it base-space}.
 In~\ref{AppendixA} we will prove the first of  eq.(\ref{eq:10-2}). The others are easy to derive.
\par
Now let us turn to the goal of this paper, i.e. to build a similarity transformation which project out of the superfield $\Phi^a$ in (\ref{eq:9-1}) its first component $\varphi^a$. Trivially it could be obtained by freezing to zero $\theta,\bar\theta$:
\[
 \left.\Phi^a\right|_{\theta,\bar\theta=0}=\varphi^a.
\]
We would like to do it in a different manner to which a physical interpretation could be associated. What we will do here is already contained in ref.\citen{ref4}. but we present it again for completeness and in order to help the reader not familiar with that reference.\\
It is easy to prove that the similarity transformation we are looking for is the following one 
\begin{equation}
  e^{-\theta\,Q-\bar Q\,\bar \theta}\,\Phi^a(t,\theta,\bar\theta)\,e^{\theta\,Q+\bar Q\,\bar \theta}=\varphi^a(t).
  \label{eq:12-1}  
\end{equation}
The proof of the formula above is contained in ref.\citen{ref4}. and for completeness we have reproduced  it  in \ref{AppendixB}. Using (\ref{eq:12-1}) it is not difficult to prove a generalisation of that formula which is the following one :
\begin{equation}
  e^{-\theta\,Q-\bar Q\,\bar \theta}\,F\left(\Phi^a(t,\theta,\bar\theta)\right)\,e^{\theta\,Q+\bar Q\,\bar \theta}=F\left(\varphi^a(t)\right)
  \label{eq:12-2}  
\end{equation}
where $F$ is a generic function.
The proof of this formula has been done in ref.\citen{ref4} and we shall not do it again here. It is also easy to prove the inverse of (\ref{eq:12-1}) and (\ref{eq:12-2}), i.e.
\begin{equation}
\Phi^a(t,\theta,\bar\theta)=
  e^{\theta\,Q+\bar Q\,\bar \theta}\varphi^a(t)\,e^{-\theta\,Q-\bar Q\,\bar \theta}
  \label{eq:13-1}  
\end{equation}
and 
\begin{equation}
F\left(\Phi^a(t,\theta,\bar\theta)\right)=
  e^{\theta\,Q+\bar Q\,\bar \theta}\,F\left(\varphi^a(t)\right)\,e^{-\theta\,Q-\bar Q\,\bar \theta}.
  \label{eq:13-2}  
\end{equation}
\par 
Note that this inverse relation cannot be obtained if we replace the operators $Q$ and $\widebar Q$ with their "hat" partners given in 
eq.(\ref{eq:10-2}). This seems to imply that the $Q$, ${\widebar Q}$ and $\widehat Q$, $\widehat{\widebar Q}$ are not completely  equivalent\footnote{ We wish to thank one referee for having brought this point to our attention.}. Anyhow in this paper we will be using only relation (\ref{eq:12-1}) and for these $Q$, $\widebar Q$  and $\widehat Q$, $\widehat{\widebar Q}$ are equivalent. We will mostly use the hat operators in the last steps of the calculations (like in  \ref{AppendixB}) because they are easier to use than the non-hat (or {\it target-space} ) operators.
\par
It is now amusing  to give a physical interpretation of the formulas above. 
We said that $\theta,\bar \theta$
 are grassmannian partners of $t$ and more details can be found in ref.\citen{ref4}. With respect to $t$ we can have the usual Heisenberg and Scroedinger picture of the system and the same should be possiible for its partners  $\theta$ and $\bar\theta$. The observables in the Heisenberg picture depend on $t$ while they don't in the Schroedinger picture. The same is possible for the grassmannian partners $\theta,\bar \theta$ of $t$. We can have a picture in which the observables  depend on $\theta,\bar\theta$ (which we shall call Heisenberg picture for $\theta,\bar\theta$) and 
one in which the observables do not depend on $\theta,\bar\theta$ (the Schroedinger picture in $\theta,\bar\theta$). Looking now at (\ref{eq:13-1}) and (\ref{eq:13-2}) we can say that the superfield $\Phi^a(t,\theta,\bar\theta)$ is the Heisenberg picture in $\theta,\bar\theta$ of the variables $\varphi^a(t)$.
 The same for the general observable $F\left(\Phi(t,\theta,\bar\theta\right))$ as proved by formula (\ref{eq:13-2}). This is the physical interpretation of the similarity transformation we mentioned at the beginning and we are amused with it !!. It may throw further light on the interplay between the superfield and its first component. For sure this throws  more light that the naive procedure of getting one from the other by naively putting $\theta,\bar\theta$ to zero.
 
\section{4-DIMENSIONAL CASE}\label{Sect3} 
In this section we will try to do the same as in Section \ref{Sect2} but for a relativistic field theory in $4$-dim.
\par We will use the Wess-Zumino model \cite{ref1.1} in the notation of ref.\citen{ref9}. The superspace in this case is  made of the usual $4$-dim. space-time $x^\mu$ and of four grassmannian variables $\left\{\theta_A\right\}|_{A=1,2}$,  $\left\{\bar\theta_{\dot A}\right\}|_{\dot A=1,2}$, where we have used the Weyl two components notation. The $\theta_A$ carries the self-representation of SL($2;\mathbb C)$, while $\bar \theta_{\dot A}$ carries the conjugate self-representation. They are all anticommuting variables:
\begin{equation}
\left[\theta_A,\theta_B\right]=\left[\bar\theta_{\dot A},\bar\theta_{\dot B}\right]=\left[\theta_A,\bar\theta_{\dot B}\right]=0, 
  \label{eq:15-1}  
\end{equation}
where $\left[(\cdot), (\cdot) \right]$ indicates the graded commutators. The indices are raised by the symbols $\epsilon^{AB}$ and $\epsilon^{\dot A\dot B}$, defined as:
\begin{equation}
\begin{matrix}
\left\{\epsilon^{AB}\right\}_{A,B=1,2}=\left(\begin{matrix}
0&1\\
-1&0
\end{matrix}\right)=\left\{\epsilon^{\dot A\dot B}\right\}_{\dot A,\dot B=1,2}\\
\left\{\epsilon_{AB}\right\}_{A,B=1,2}=\left(\begin{matrix}
0&-1\\
1&0
\end{matrix}\right)=\left\{\epsilon_{\dot A\dot B}\right\}_{\dot A,\dot B=1,2}.\\
\end{matrix}
  \label{eq:16-1}  
\end{equation}
So the objects with the indices up are defined as:
\begin{equation*}
\psi^A\equiv \epsilon^{AB}\,\psi_B=-\psi_B\,\epsilon^{BA},\quad \bar \psi^{\bar A}\equiv \epsilon^{\dot A\dot B}\bar \psi_{\dot B}.
\end{equation*}
It is then easy to obtain the following relations:
\begin{equation}
\begin{matrix}
\psi_A&=&\epsilon_{AB}\,\psi^B&=&-\psi^B\,\epsilon_{BA}\\
\bar\psi_{\dot A}&=&\epsilon_{\dot A\dot B}\,\bar\psi^{\dot B}&=&-\bar\psi^{\dot B}\,\epsilon_{\dot B\dot A}.
\end{matrix}
  \label{eq:16-2}  
\end{equation}
 In performing the scalar products among Weyl spinors we will use the convention of summing upper left indices with lower right indices and lower left dotted indices with upper right dotted indices: 
\begin{equation}
\left(\psi \,\chi\right)\equiv \psi^A\,\chi_A,\quad \left(\bar\psi \,\bar\chi\right)\equiv \bar\psi_{\dot A}\,\bar\chi^{\dot A}.
  \label{eq:16-3}  
\end{equation} 
In the superspace $\{x^{\mu};\theta^{\alpha},\bar \theta_{\dot \alpha}\}$ we can define a scalar superfiel as follows
\begin{equation}
\begin{matrix}
\Phi(x;\theta,\bar\theta)&=&f(x)+\theta\,\phi(x)+\bar\theta\,\bar\chi(x)+(\theta\theta)\,m(x)\\
&+& (\bar\theta\bar\theta)\,n(x)+\left(\theta\sigma^{\mu}\bar\theta\right)\,V_{\mu}(x)+(\theta\theta)\bar\theta\,\bar\lambda\\
&+&(\bar\theta\bar\theta)\theta\,\psi+(\theta\theta)(\bar\theta\bar\theta)\,d(x),\quad\quad\quad\quad\quad\,
\end{matrix}
  \label{eq:17-1}  
\end{equation}
where $f(x),m(x),n(x),d(x)$ are scalar and pseudoscalar complex fields, $\psi^\alpha(x),\,\phi^\alpha(x)$ are spinor fields which transform according to the self-representation of SL$(2;\mathbb C)$ (left-handed Weyl spinors), $\bar\chi(x)^{\dot \alpha},\,\bar \lambda(x)^{\dot \alpha}$ are  spinor fields which transform according to the conjugate self-representation of SL$(2;\mathbb C)$ (right-handed Weyl spinors)
and $V_\mu$ is a Lorentz vector field.  
In what follows we will use  the notation $\sigma^\mu=(\mathbb{I},\sigma^i)$, where $\sigma^i$ are the Pauli matrices. Notice that the $\sigma^\mu$ carry mixed indices $(\sigma^\mu)_{A\dot A}$ while its adjoint is defined as
\[
\left(\bar\sigma^\mu\right)^{\dot AA}\equiv\epsilon^{AB}\epsilon^{\dot A\dot B} \sigma_{B\dot B}^\mu. 
\]  
The various expressions without explicit index notation appearing in (\ref{eq:17-1}) are defined as follows: 
\begin{equation}
\begin{matrix}
\theta\,\phi=\theta^\alpha\,\phi_\alpha\quad\quad\quad\quad\quad\quad\\
\bar\theta\,\bar\chi=\bar\theta_{\dot \alpha}\,\bar\chi^{\dot \alpha}\quad\quad\quad\quad\quad\quad\\
(\theta\sigma^\mu\bar\theta)\,V_\mu=(\theta^\alpha\,\sigma^\mu_{\alpha\dot\alpha}\,\bar\theta^{\dot \alpha})\,V_\mu\\
\bar\theta\bar\lambda=\bar\theta_{\dot \alpha}\bar\lambda^{\dot \alpha}\quad\quad\quad\quad\quad\quad\,\,\\
\theta\psi=\theta^\alpha\psi_\alpha\quad\quad\quad\quad\quad\quad\,\,
\end{matrix}.
\end{equation}
\par Note that differently than the scalar superfiend in 1-dim of eq.(\ref{eq:9-1}), the scalar superfield in 4-dim of eq.(\ref{eq:17-1}) contains up to 4 powers in $\theta,\,\bar\theta$, because these $\theta$ and $\bar \theta$ carry two indices. This point will be crucial in calculations we will perform later on.
\par In the Weyl-formulation the supersymmetry charges $Q_H,\,\bar Q_H$ are two  and each carries two indices:
\[
\begin{matrix}
(Q_H)_A,&\,&A=1,2\\
(\bar Q_H)_{\dot A},&\,&\dot A=1,2.
\end{matrix}
\]
 The supersymmetry (or Super-Poincar\'e algebra) is the following one
\begin{equation}
\begin{matrix}
\left[(Q_H)_A,(Q_H)_B\right]&=&0\\
\left[(\bar Q_H)_{\dot A},(\bar Q_H)_{\dot B}\right]&=&0\\
\left[(Q_H)_A,(\bar Q_H)_{\dot B}\right]&=&2\,\left(\sigma^\mu\right)_{A\dot B}\,P_\mu
\end{matrix},
  \label{eq:18-1}  
\end{equation} 
 where $P_\mu$ is the 4-momentum operator.
 The representation as differential operators in super-space of the supersymmetry operators (\ref{eq:18-1}) is the following \cite{ref9} one:
\begin{equation}
\begin{matrix}
\left(\widehat{Q}_H\right)_A&=&-i\left(\partial_A-\sigma^\mu_{A\dot B}\,\bar\theta^{\dot B}\,\partial_\mu\right)\\
\left(\widehat{\widebar Q}_H\right)^{\dot A}&=&-i\left(\bar\partial^{\dot A}-i\,\left(\bar\sigma^\mu\right)^{\dot AB}\,\theta_{\dot B}\,\partial_\mu\right)
\end{matrix}
  \label{eq:19-1}  
\end{equation} 
 where 
 \[
\partial_A\equiv\frac{\partial}{\partial\theta^A},\quad\bar\partial^{\dot A}\equiv\frac{\bar \partial}{\partial\bar\theta_{\dot A}},\quad\partial_\mu\equiv\frac{\partial}{\partial x^\mu}. 
 \]
 From now on , in analogy with the 1-dim.case, let us indicate with $\widehat Q$ and $\widehat{\widebar Q}$ the generators:
 
 \begin{equation} 
 \begin{matrix} 
 \widehat Q_A&\equiv&+i\,\partial _A\\ \widehat{\widebar Q}^{\dot A}&\equiv&+i\,\bar\partial^{\dot A}. \end{matrix}
 \label{eq:19-2}
 \end{equation}

 
Note that differently than  $\widehat Q_H$ and $\widehat{\widebar Q}_H$, the $\widehat{Q}$ and $\widehat{\widebar Q}$ anticommute
\begin{equation}
\left[\widehat Q_A,\widehat{\widebar Q}^{\dot B}\right]=0
\label{eq:19-3}
\end{equation}
and the same happens with the $\widehat{Q}_A$ among themselves and the $\widehat{\widebar Q }^{\dot B}$ as well. 
Via the analog of eq.(\ref{eqinserzione:8}) we can build the target-space counterpart of $\widehat{Q}_A,\,\widehat{\widebar Q }^{\dot B}$ which we shall indicate without "\,$\widehat{\,\,\,}$\," as $Q_A$ and $\widebar Q^{\dot B}$. 
Following what we did in the 1-dim case, the next step consists in  performing  the following similarity transformation:
\begin{equation}
e^{-\theta\,{Q}-{\widebar Q}\,\bar \theta}\,\Phi\, e^{\theta\,{Q}+{\widebar Q}\,\bar \theta}
\label{eq:20-1}
\end{equation}
on the superfield (\ref{eq:17-1}) in 4-dim. Let us stress again  that ${Q},\,{\widebar Q}$  are not the supersymmetry charges but the generators of translations in $\theta$ and $\bar \theta$ in analogy to  the 1-dim case. In this manner, if (\ref{eq:20-1}) would give the first component of the superfield, we could say that $\Phi$ is the Heisenberg picture representation of its first component. 
\par 
In order to perform the calculation (\ref{eq:20-1}) we need first to find out the expression for the exponentials surrounding $\Phi$. It is easy to prove that :
\begin{equation}
\begin{matrix}
e^{-\theta\,{Q}-{\widebar Q}\,\bar \theta}&=&\left(1-\theta\,{Q}+\cfrac{1}{2}\,\left(\theta \,Q\right)^2\right)\,\left(1-\bar\theta\,{\widebar Q}+\cfrac{1}{2}\,\left(\bar\theta \,\widebar Q\right)^2\right)\quad\quad\quad\quad\,\,\\
&=&\left(1-\theta\,{Q}-\bar\theta\,{\widebar Q}-\theta\bar\theta\,{Q}{\widebar Q}-\cfrac{1}{4}\,(\theta\theta)\left({Q}{Q}\right)-\cfrac{1}{4}\,(\bar\theta\bar\theta)\left({\widebar Q}{\widebar Q}\right)\right.\\
&+&\left.\cfrac{1}{4}\,(\theta\theta)\bar \theta\left({\widebar Q}{\widebar Q}\right){\widebar Q}+
\cfrac{1}{4}\,\theta\,(\bar\theta\bar\theta){Q}\left({\widebar Q}{\widebar Q}\right)\right.\\
&+&\left.\cfrac{1}{16}\,(\theta\theta)\,(\bar\theta\bar\theta)\left({Q}{Q}\right)\left({\widebar Q}{\widebar Q}
\right)\right).
\end{matrix}.
\label{eq:21-1}
\end{equation}
The other exponential $e^{\theta\,{Q}+{\widebar Q}\,\bar \theta}$ is the same as  the R.H.S. of (\ref{eq:21-1}) with the replacement  $\theta\to -\theta$ and $\bar\theta\to -\bar\theta$. We note that, because $\theta_\alpha$ carries two indices,  in the R.H.S. of (\ref{eq:21-1}) we have up to quartic terms like $(\theta\theta)(\bar\theta\bar\theta)$. This is the crucial difference with respect to the 1-dim case. In (\ref{eq:21-1}) the convention we have used is to put round brackets $(\,\,)$ around quantities which have their indices saturated with each other like it is done in (\ref{eq:16-3}). Moreover in (\ref{eq:21-1}) we use the convention of putting in front all the strings of $\theta$ and $\bar\theta$ and behind the operators $ Q$ and ${\widebar Q}$.
\par  Let us now go back to (\ref{eq:21-1}): in order to understand how the indices are saturated among quantities which are not incapsulated inside round brackets, let us explicitly write few terms present in (\ref{eq:21-1}) :
\begin{equation}
\begin{matrix}
\theta\bar\theta\,{Q}{\widebar Q}&\equiv& \theta^\alpha\bar \theta_{\dot \beta}\,{Q}_\alpha{\widebar Q}^{\dot\beta}\\
(\theta\theta)\bar\theta\,\left({Q}{Q}\right){\widebar Q}&\equiv&(\theta\theta)\bar\theta_{\dot \alpha} \,\left({Q}{Q}\right){\widebar Q}^{\dot\alpha}\\
\theta(\bar\theta\bar\theta)\, Q\left({\widebar Q}{\widebar Q}\right)&\equiv&\theta^\alpha\left(\bar\theta\bar\theta\right)\,{Q}_\alpha\left({\widebar Q}{\widebar Q}\right).
\end{matrix}
\label{eq:22-1}
\end{equation}
Let us now insert (\ref{eq:21-1}) and its inverse into eq. (\ref{eq:20-1}) and let us perform the calculations. These are rather long  and some samples of them are reproduced in \ref{AppendixC}. 
Looking at this appendix we want to point out for the reader that, once we get to (\ref{eq:C-1}), its R.H.S. can be calculated in the following manner. First we use recoursevly the analog of eq.(\ref{eqinserzione:8}) and turn all commutators in actions of the derivative operators on the superfield, second we perform these simple derivations like we did in the last steps of  \ref{AppendixB}. The final result is the following: 
\begin{equation}
e^{-\theta\,{Q}-{\widebar Q}\,\bar \theta}\,\Phi\, e^{\theta\,{Q}+{\widebar Q}\,\bar \theta}=f(x)-2\,(\theta\theta)\,m(x)-2\,(\bar\theta\bar\theta)\,n(x)+4\,(\theta\theta)(\bar\theta\bar\theta)\,d(x),
\label{eq:23-1}
\end{equation}
where $f(x),\,m(x),\,n(x),\,d(x)$ are the scalar and pseudocalar fields component of the superfield (\ref{eq:17-1}).
\par From eq.(\ref{eq:23-1}) we  notice unfortunatly that our similarity transformation does not give us  the first component $f(x)$ of the superfield like in the 1-dim. case  (see (\ref{eq:12-1})). So this is not the transformation we are looking for. Anyhow the result (\ref{eq:23-1}) is consistent with (\ref{eq:12-1}). In fact if the $\theta$ and $\bar\theta$ in (\ref{eq:23-1}) had no index, then $\theta\theta=0=\bar\theta\bar\theta$ and the last three terms on the R.H.S. of (\ref{eq:23-1}) would be zero leaving only the first term like in (\ref{eq:12-1}). 
\par  At this point we were tempted by the idea that if we had started with a superfield containing less components than the one in (\ref{eq:17-1}), we may have ended up in the R.H.S. of (\ref{eq:23-1}) with less fields or even just one. The superfield with less components that we know are the chiral and antichiral superfields.They are also called lefthanded (L) or righthanded (R) chiral superfields. We will use the notation of ref.\citen{ref9} and they are given by:
\begin{equation}
\begin{matrix}
\Phi_L&=&A(x)+\sqrt 2\,\theta\,\psi(x)+i\,\left(\theta\sigma^\mu\bar\theta\right)\partial_\mu A(x)+(\theta\theta)\,F(x)\\
&+&\sqrt 2\,i\,\left(\theta\sigma^\mu\bar\theta\right)\partial_\mu\left(\theta\,\psi\right)-\cfrac{1}{4}\,(\theta\theta)(\bar\theta\bar\theta) \,\Box \,A(x)\\
&&\\
\Phi_R&=&A^{*}(x)+\sqrt 2\,\bar\theta\,\bar\psi(x)-i\,\left(\theta\sigma^\mu\bar\theta\right)\partial_\mu A^*(x)+(\bar\theta\bar\theta)\,F^{*}(x)\\
&-&\sqrt 2\,i\,\left(\theta\sigma^\mu\bar\theta\right)\partial_\mu\left(\bar\theta\,\bar\psi\right)-\cfrac{1}{4}\,(\theta\theta)(\bar\theta\bar\theta) \,\Box \,A^*(x),\\
\label{eq:24-1}
\end{matrix}
\end{equation}
where $A$ and $F$ are  complex scalar fields of no definite parity and $\psi$ is a Weyl fermion field. Doing the similarity transformation analog of (\ref{eq:20-1}) for $\Phi_L$ and $\Phi_R$ we get: 
\begin{equation}
\begin{matrix}
e^{-\theta\,{Q}-{\widebar Q}\,\bar \theta}\,\Phi_R\, e^{\theta\,{Q}+{\widebar Q}\,\bar \theta}&=&A^*(x)-2\,(\theta\theta)\,F^*(x)-(\theta\theta)(\bar\theta\bar\theta)\,\Box\,A^*(x),\\
e^{-\theta\,{Q}-{\widebar Q}\,\bar \theta}\,\Phi_L\, e^{\theta\,{Q}+{\widebar Q}\,\bar \theta}&=&A(x)-2\,(\theta\theta)\,F(x)-(\theta\theta)(\bar\theta\bar\theta)\,\Box\,A(x).
\end{matrix}
\label{eq:25-1}
\end{equation}
So we conclude that also for the chiral superfields our similarity transformation does not lead us to the first component but to a combination of the scalar fields present in the original superfield.
The calculations that lead to the results (\ref{eq:25-1}) are rather easy. Basically we started identifying the fields present on the R.H.S. of (\ref{eq:17-1}) with the analog ones present on the R.H.S. of (\ref{eq:24-1}). Some of them turned out to be zero. Once this identification was done we plugged it in the R.H.S. of (\ref{eq:23-1}) and immediately we got eqs.(\ref{eq:25-1}). \par
The reader may be tempted to explore another route  that is the one of using linear combinations of $Q,\widebar{Q},Q_H,\widebar{Q}_H$ instead of just $Q,\,\widebar Q$. Also this route does not work because these combinations can only get rid of terms linear in $\theta$ or in $\bar \theta$ and not quadratic terms like those which survive on the R.H.S. of eq.(\ref{eq:23-1}). In the next section we will attack this problem of the quadratic terms. 
\section{PLAYING WITH THE ENVELOPING ALGEBRA}\label{Sect4} 
Let us  go back  to (\ref{eq:23-1}) and let us define with $\widetilde \Phi(x,\theta,\bar\theta)$ the R.H.S. of (\ref{eq:23-1}):
\begin{equation}
\widetilde \Phi\equiv f(x)-2\,(\theta\theta)\,m(x)-2\,(\bar\theta\bar\theta)\,n(x)+4\,(\theta\theta)(\bar\theta\bar\theta)\,d(x).
\label{eq:27-1}
\end{equation}
So (\ref{eq:23-1}) can be written as: 
\begin{equation}
e^{-\theta\,{Q}-{\widebar Q}\,\bar \theta}\,\Phi\, e^{\theta\,{Q}+{\widebar Q}\,\bar \theta}=\widetilde \Phi.
\label{eq:27-2}
\end{equation}
 As we said in the previous section the reason that the R.H.S. of (\ref{eq:27-1}) is different from $f(x)$ is because the grassmannian variables $\theta$ are two: $\theta^A,\,A=1,2$, and the same for the $\bar \theta$ and so terms quadratic in $\theta$ and $\bar \theta$ are not identically zero like it happened instead in $D=1$.   
 We should then look for an operator whose base-space differential counterpart kills terms like $(\theta\theta)$ and $(\bar\theta\bar\theta)$
present on the R.H.S. of eq.(\ref{eq:27-1}). 
The base-space differential counterpart of $Q$ is the $\widehat{Q}$ of eq.(\ref{eq:19-2}) and it annihilates only one power of $\theta$.
We need something like a double derivative, we can try on $\widetilde\Phi$ the following similarity transformation 
\begin{equation}
e^{-\alpha\,(\theta\theta)\,({Q}{Q})-\beta\,(\bar\theta\bar\theta)\,({\widebar Q}{\widebar Q})}\,\widetilde\Phi\, e^{\alpha\,(\theta\theta)\,({Q}{Q})+\beta\,(\bar\theta\bar\theta)\,({\widebar Q}{\widebar Q})},
\label{eq:28-1}
\end{equation}
where $(QQ)$ is the target-space version of a double derivative. The problem here is that $\widetilde\Phi$ is not a scalar field and we cannot use eq.(\ref{eqinserzione:8}) in order to turn the target-field operators $Q,\widebar{Q}$ into the derivative operators $\widehat Q,\widehat{\widebar Q}$. To solve this problem let us indicate with $\widetilde Q,\widetilde{\widebar Q}$ some abstract operators which would act as translation operators in $\theta$ and $\bar\theta$. They would be the analog of the abstract translation operators $\hat P$ which on a function $F(x)$ of $x$ acts as follows:
\[
e^{-i\hat Pa}\,F(x)\, e^{i\hat Pa}=F(x+a).
\]
So eq.(\ref{eq:28-1}) will be replaced by the following one: 
\begin{equation}
e^{-\alpha\,(\theta\theta)\,(\widetilde{Q}\widetilde{Q})-\beta\,(\bar\theta\bar\theta)\,(\widetilde{\widebar Q}\widetilde{\widebar Q})}\,\widetilde\Phi\, e^{\alpha\,(\theta\theta)\,(\widetilde{Q}\widetilde{Q})+\beta\,(\bar\theta\bar\theta)\,(\widetilde{\widebar Q}\widetilde{\widebar Q})},
\label{eq:28-1bis}
\end{equation}
where $\alpha$ and $\beta$ are parameters to be adjusted in order for the result of (\ref{eq:28-1bis}) to be the $f(x)$ of (\ref{eq:27-1}). If this were possible then the combination of the similarity transformation (\ref{eq:27-2}) and (\ref{eq:28-1}) would manage to turn the superfield into its first component which was our original goal. A long set of calculations (of which some samples are reported in \ref{AppendixD}) shows that a choice of $\alpha,\,\beta$ exists, which leads to the result we want and it is $\alpha=\beta=1/4$.
Inserting these values we can say that we have  basically proved the following:
\begin{equation}
e^{\frac{1}{4}\,\left[-\,(\theta\theta)\,(\widehat{Q}\widehat{Q})-\,(\bar\theta\bar\theta)\,(\widehat{\widebar Q}\widehat{\widebar Q})\right]}\,\widetilde\Phi\, e^{\frac{1}{4}\,\left[\,(\theta\theta)\,(\widehat{Q}\widehat{Q})+\,(\bar\theta\bar\theta)\,(\widehat{\widebar Q}\widehat{\widebar Q})\right]}=f(x).
\label{eq:29-1}
\end{equation}
In the formula above we have replaced the $\widetilde Q$ with $\widehat{Q}$ because their action is the same. Note that the operators which we used are $\widehat{Q}\widehat{Q}$ and $\widehat{\widebar Q}\widehat{\widebar Q}$  which do not belong to the algebra of (\ref{eq:19-1}) and (\ref{eq:19-2}) but to their enveloping algebra. 
In general {\it  "enveloping algebras"} are defined as an  extension of Lie algebras which contain, beside the operation of commutation, also the one of multiplication of operators. So enveloping algebras contain the basic operators of the Lie algebras plus all of its higher powers and linear combinations of them. \par Let us now summarize what we did and combine  into a single operation the transformation (\ref{eq:23-1}) which brings us from $\Phi$ to $\widetilde \Phi$ with the (\ref{eq:29-1}) which carries  us from $\widetilde \Phi$ to $f(x)$ :
\begin{equation}
e^{\left[-\frac{1}{4}\,(\theta\theta)\,(\widehat{Q}\widehat{Q})-\frac{1}{4}\,(\bar\theta\bar\theta)\,(\widehat{\widebar Q}\widehat{\widebar Q})-\theta\widehat{Q}-\bar\theta\widehat{\widebar Q}\right]}\,\Phi\, e^{\left[\frac{1}{4}\,(\theta\theta)\,(\widehat{Q}\widehat{Q})+\frac{1}{4}\,(\bar\theta\bar\theta)\,(\widehat{\widebar Q}\widehat{\widebar Q})+\theta\widehat{Q}+\bar\theta\widehat{\widebar Q}\right]}=f(x).
\label{eq:30-1}
\end{equation}
We have written above  all over the "hat" operators because by now it should be clear that its action is the same as the "non-hat" operators if one uses the relation (\ref{eqinserzione:8}).
The combined transformation above does not seem to be the Heisenberg picture transformation 
of the form (\ref{eq:12-1}) . One last attempt we can do is to modify the $\widehat Q, \widehat{\widebar Q},\theta,\bar\theta$ in order to reduce (\ref{eq:30-1}) to the form (\ref{eq:12-1}). Let us look at the exponential on the left of $\Phi$ in  (\ref{eq:30-1}) and expand it in $\theta,\bar\theta$:
\begin{equation}
\begin{matrix}
e^{\left[-\frac{1}{4}\,(\theta\theta)\,(\widehat{Q}\widehat{Q})-\frac{1}{4}\,(\bar\theta\bar\theta)\,(\widehat{\widebar Q}\widehat{\widebar Q})-\theta\widehat{Q}-\bar\theta\widehat{\widebar Q}\right]}&=&\\
&=&1-\theta\widehat{Q}-\left(\bar\theta\widehat{\widebar Q}\right)-\frac{1}{2}\,(\theta\theta)\left(\widehat{Q}\widehat{Q}\right)-(\theta\bar\theta)\,\left(\widehat{Q}\widehat{\widebar Q}\right)\\
&-&\frac{1}{2}\,(\bar\theta\bar\theta)\left(\widehat{\widebar Q}\widehat{\widebar Q}\right)+\frac{1}{2}\,(\theta\theta)\bar\theta\left(\widehat{Q}\widehat{Q}\right)\widehat{\widebar Q}\\&+&\frac{1}{2}\,(\bar\theta\bar\theta)\left(\widehat{\widebar Q}\widehat{\widebar Q}\right)\left(\theta\widehat{Q}\right)+\frac{1}{4}\,(\theta\theta)(\bar\theta\bar\theta)\left(\widehat{Q}\widehat{Q}\right)\left(\widehat{\widebar Q}\widehat{\widebar Q}\right).
\end{matrix}
\label{eq:30-2}
\end{equation}
This should be compared with the first exponential in (\ref{eq:12-1}) (but with $\theta$ and $\bar\theta$ carrying  indices like  in (\ref{eq:21-1})).
Note that in (\ref{eq:30-2}) the coefficients of the fourth, the sixth, seventh and eight terms are different than those in (\ref{eq:21-1}). We may try to fix things by rescaling $\theta$ and $\bar\theta$ but, as the second and third terms  are the same in (\ref{eq:30-2}) and (\ref{eq:21-1}), we would have to rescale also the $\widehat{Q}$ and $\widehat{\widebar Q}$ in an inverse manner. This rescaling unfortunately leaves invariant also the fourth term in  (\ref{eq:30-2}) which instead has to be changed to match the analogous one in (\ref{eq:21-1}). So a rescaling does not work. We tried other kind of transformations like: 
\begin{equation}
\widehat{Q}_\alpha\to\widehat{Q}_\alpha+\theta_\alpha\left(\widehat{Q}^\beta\widehat{Q}_\beta\right).
\label{eq:32-1}
\end{equation} 
This new charge generates still a translation in $\theta_\alpha$ like the $\widehat{Q}_\alpha$ but it does not act as a translation on quadratic function of $\theta$, so its geometric meaning is lost. 

\section{CONCLUSIONS}
\par In this paper we have built the similarity transformation which reduces a superfield to its first component. While in the 1-dim. this transformation can be built from the generators of the supersymmetry algebra (or better from a supertranslation), in 4-dim this was not possible and we had to use the generators of the enveloping algebra. This took away the possibility of  interpreting the similarity transformation as an Heisenberg picture transformation in the grassmannian time like in 1-dim.
\par Some extra work  could  be done along the directions indicated at the end of Section \ref{Sect4}. In particular we should try to see if there are changes on the generators and the space $\theta,\bar\theta$ which would turn our similarity transformation in some translation in a modified space. The few attempts we tried do not seem to work but there may be other more interesting ones to explore. For sure they will be  transformations with generators on the enveloping algebra analogous to those in  (\ref{eq:32-1}). In general we feel that the {\it enveloping algebra} is a space which deserves to be explored more.

\appendix
\section{}\label{AppendixA}
In this appendix we will check  relation (\ref{eqinserzione:8}) with the choice of operators given in the first of eqs.(\ref{eq:8-3})  and  (\citen{ref5.2}), i.e.:
\begin{equation}
\left[ Q,\Phi\right]=-\widehat Q\,\Phi
\label{eq:A-1}
\end{equation}
Using eq.(\ref{eq:10-2}), the R.H.S. of (\ref{eq:A-1}) turns out to be  
\begin{equation*}
\begin{matrix}
-\widehat Q\,\Phi&=&\partial_\theta\left[\varphi^a+\theta\,c^a+\bar\theta\,\omega^{ab}\,\bar c_b+i\,\bar\theta\theta\,\omega^{ab}\lambda_b\right]\\
&=&c^a-i\,\bar\theta\,\omega^{ab}\lambda_b.
\end{matrix}
\end{equation*}
while the L.H.S. is 
\begin{equation*}
\begin{matrix}
\left[i\,\lambda_d\,c^d,\varphi^a+\theta\,c^a+\bar\theta\,\omega^{ab}\,\bar c_b+i\,\bar\theta\theta\,\omega^{ab}\lambda_b\right]&=&\\
&=&\left[i\,\lambda_d\,c^d,\varphi^a+i,\bar\theta,\omega^{ab}\bar c_b\right]\\
&=&i\,\left[\lambda_d,\varphi^a\right]\,c^d-i\,\lambda_d\left[c^d,\bar c_b\right]\bar\theta\,\omega^{ab}\\
&=&i\,(-i)\delta^a_d\,c^d-i\lambda_d\,\delta^d_b \,\bar\theta\, \omega^{ab}\\
&=&c^a-i\,\bar\theta\,\omega^{ab}\lambda_b,
\end{matrix}
\end{equation*}
where we have used the commutation relations (\ref{eq:8-1}). So the R.H.S.. and L.H.S. of (\ref{eq:A-1}) are equal and this concludes our proof.
\section{}\label{AppendixB}
In this appendix we will reproduce formula (\ref{eq:12-1}):
\begin{equation}
e^{-\theta\,Q-\bar Q\,\bar \theta}\,\Phi^a(t,\theta,\bar\theta)\,e^{\theta\,Q+\bar Q\,\bar \theta}=\varphi^a(t).
\label{eq:B-1}
\end{equation}
The L.H.S. of the previous equation, because of the anticommuting character of $\theta$ and  $\bar\theta$, is equal to:  
\begin{equation}
\left(1-\theta Q\right)\left(1-\widebar Q\bar{\theta}\right)\Phi^A\left(1+\theta Q\right)\left(1+\widebar Q\bar{\theta}\right)
\label{eq:B-2}
\end{equation}
and performing the products contained in (\ref{eq:B-2}) we get that it is equal to: 
\begin{equation}
\Phi^a-\theta\left[Q,\Phi^a\right]+\bar \theta\left[\widebar Q,\Phi^a\right]+\bar\theta\theta\left[\left[Q,\Phi^a\right],\widebar Q\right],
\label{eq:B-3}
\end{equation}
where the commutators are intended as graded commutators.
We can perform the calculations of the various commutators in (\ref{eq:B-3}) using the expressions of $Q$ and $\bar Q$ from eq.(\citen{ref2.1}) and the commutators of the basic variables given in eq.(\citen{ref1.2}) but we will obtain the same result if we use eq.(\ref{eqinserzione:8}) and the differential operator version of $Q$ and $\widebar Q$. These last calculations are easier to do. The relation (\ref{eq:B-3}) can then be written as:
\begin{equation}
\Phi^a-\theta\,\partial_\theta\Phi^a-\bar\theta\,\partial_{\bar\theta}\Phi^a-\bar\theta\theta\,\partial_{\bar\theta}\partial_\theta\Phi^a.
\label{eq:B-4}
\end{equation}
It is easy to perform in (\ref{eq:B-4}) the derivative with  respect to $\theta$ and $\bar\theta$ of the superfields $\Phi^a$ using its expression (\ref{eq:9-1}). The result turns out to be: 
\begin{equation}
\Phi^a-\theta\,\partial_\theta\Phi^a-\bar\theta\,\partial_{\bar\theta}\Phi^a-\bar\theta\theta\,\partial_{\bar\theta}\partial_\theta\Phi^a=\varphi^a.
\end{equation}
This proves relation (\ref{eq:B-1}).
\section{}\label{AppendixC}
In this appendix we will report a sample of the calculations which lead to eq.(\ref{eq:23-1}). It is hard  to report all of them because they are extremely long. Those which lead to  (\ref{eq:21-1}) are relatively easy and short and we will not present them. Inserting the  (\ref{eq:21-1}) and its inverse  into  (\ref{eq:20-1}) we get: 
\begin{equation}
\begin{matrix}
e^{-\theta\,{Q}-\bar \theta\,{\widebar Q}}\,\Phi\, e^{\theta\,{Q}+\bar \theta{\widebar Q}}&=&\Phi-\theta\,\left[{Q},\Phi\right]-\bar\theta\,\left[{\widebar Q},\Phi\right]\\
&+&\cfrac{(\theta\theta)}{4}\,\left[\left[{Q},\Phi\right],{Q}\right]+\cfrac{(\bar\theta\bar\theta)}{4}\,\left[\left[{\widebar Q},\Phi\right], {\widebar Q}\right]\\
&+&\theta\bar\theta\,\left[\left[{Q},\Phi\right],{\widebar Q}\right]+\cfrac{(\theta\theta)\bar\theta}{4}\left[\left[\left[{Q},\Phi\right],\widehat{Q}\right],{\widebar Q}\right]\\
&+&\cfrac{\theta(\bar\theta\bar\theta)}{4}\left[\left[\left[{Q},\Phi\right],{\widebar Q}\right],{\widebar{Q}}\right]
-\cfrac{(\theta\theta)(\bar\theta\bar\theta)}{16}\left[\left[\left[\left[{Q},\Phi\right],{Q}\right],{\widebar{Q}}\right],{\widebar{Q}}\right].
\label{eq:C-1}
\end{matrix}
\end{equation}
These commutators can be easily performed by using the four-dimensional analog of eq.(\ref{eqinserzione:8}) which turns the commutators into actions of the derivative operators $\widehat Q,\widehat{\widebar Q}$ on $\Phi$ and the result is eq.(\ref{eq:23-1}), 
\par Going now back to eq.(\ref{eq:C-1}) let us show how we got, for example,  the seventh term on the R.H.S. i.e.:
\[\cfrac{(\theta\theta)\bar\theta}{4}\left[\left[\left[{Q},\Phi\right],{Q}\right],{\widebar Q}\right].\] 
Consider that in (\ref{eq:20-1}), inserting (\ref{eq:21-1}) and its inverse and performing the products, the terms which are proportional to $\theta\theta\bar\theta$ are given by
\begin{equation}
\begin{matrix}
-\cfrac{1}{4}\,(\theta\theta\bar\theta)\Phi\left({Q}{Q}\right){\widebar Q}+\left(\theta{Q}\right)\Phi\left(\theta\bar\theta\,{Q}{\widebar Q}\right)+\cfrac{1}{4}\,\left(\bar\theta{\widebar Q}\right)\Phi(\theta\theta)\left({Q}{Q}\right)\\
-\cfrac{1}{4}\,(\theta\theta)\left({Q}{Q}\right)\Phi\left(\bar\theta{\widebar Q}\right)+\cfrac{1}{4}\,(\theta\theta)\bar\theta\left({Q}{Q}\right){\widebar Q}\Phi-\theta\bar\theta\,{Q}{\widebar Q}\,\Phi\left(\theta{Q}\right).\quad
\label{eq:C-2-1}
\end{matrix}
\end{equation}
 The first term in (\ref{eq:C-2-1}) can be written as 
 \begin{equation}
 -\cfrac{1}{4}\left(\theta^A\theta_A\,\bar\theta_{\dot A}\,\Phi\,{Q}^B{Q}_B{\widebar Q}^{\dot A}\right)=-\cfrac{1}{4}\left((\theta\theta)\left(\bar\theta\,\Phi\left({Q}{Q}\right){\widebar Q}\right)\right).
\label{eq:C-2-2}
 \end{equation}
The second term, using the identity
\[
\theta^A\theta^B=-\frac{1}{2}\epsilon^{AB}(\theta\theta),
\]
can be turned into 
\begin{equation}
\begin{matrix}
\theta^A\,{Q}_A\Phi\,\theta^B\bar\theta_{\dot A}\,{Q}_B{\widebar Q}^{\dot A}&=&\theta^A\theta^B\bar\theta_{\dot A}\,{Q}_A\Phi\,{Q}_B{\widebar Q}^{\dot A}\\
&=&-\cfrac{1}{2}\,\epsilon^{AB}(\theta\theta)\bar\theta_{\dot A}\,{Q}_A\Phi\,{Q}_B{\widebar Q}^{\dot A}\\
&=&\cfrac{1}{2}\,(\theta\theta)\bar\theta\left({Q}\,\Phi\,{Q}\right){\widebar Q},\quad
\label{eq:C-2-3}
\end{matrix}
\end{equation}
where 
\[
\left({Q}\,\Phi\,{Q}\right)={Q}^A\Phi\,{Q}_A.
\]
In the same manner the third term in (\ref{eq:C-2-1}) can be re-written as:
\begin{equation}
\frac{1}{4}\,\left(\bar\theta{\widebar Q}\right)\Phi(\theta\theta)\left({Q}{Q}\right)=\frac{1}{4}\,(\theta\theta)\,\left(\bar\theta\,{\widebar Q}\Phi\left({Q}{Q}\right)\right)
\label{eq:C-3-1}
\end{equation}
and the same for the fourth term:
\begin{equation}
-\cfrac{1}{4}\,(\theta\theta)\left({Q}{Q}\right)\Phi\left(\bar\theta{\widebar Q}\right)=-\frac{1}{4}\,(\theta\theta)\left(\bar\theta\left({Q}{Q}\right)\Phi\,{\widebar Q}\right).
\label{eq:C-3-2}
\end{equation}
The sixth term of (\ref{eq:C-2-1}) gives the following expression:
\begin{equation}
\begin{matrix}
-\theta\bar\theta\,{Q}{\widebar Q}\,\Phi\left(\theta{Q}\right)&=&-\theta^A\bar\theta_{\dot A}\,{Q}_A
{\widebar Q}^{\dot A}\Phi\left(\theta^B{Q}_B\right)\\
&=&\theta^A\theta^B\bar\theta_{\dot A}\,{Q}_A{\widebar Q}^{\dot A}\Phi\,Q_B\\&=&\frac{1}{2}\,(\theta\theta)\left(\bar{\theta}\,{Q}{\widebar Q}\Phi\,{Q}\right),
\end{matrix}
\label{eq:C-3-3}
\end{equation}
where again we have used the identity $\theta^A\theta^B=-\frac{1}{2}\epsilon^{AB}(\theta\theta)$.
\par The fifth term is : 
\begin{equation}
\frac{1}{4}\,(\theta\theta)\bar\theta\left({Q}{Q}\right){\widebar Q}\Phi.
\label{eq:C-3-4}
\end{equation}
Summing up (\ref{eq:C-2-2}),(\ref{eq:C-2-3}),(\ref{eq:C-3-1}),(\ref{eq:C-3-2}),(\ref{eq:C-3-3})
we get : 
\begin{equation}
\begin{matrix}
\cfrac{1}{4}\,(\theta\theta)\,\bar\theta_{\dot A}\left[-\Phi\left({Q}{Q}\right){\widebar Q}^{\dot A}+2\,\left({Q}\Phi{Q}\right){\widebar Q}^{\dot A}+{\widebar Q}^{\dot A}\Phi\left({Q}{Q}\right)\right.\\
&&\\
\left.-\left({Q}{Q}\right)\Phi{\widebar Q}^{\dot A}+2\,{Q^A}{\widebar Q}^{\dot A}\Phi\,{Q}_A+\left({Q}{Q}\right){\widebar Q}^{\dot A}\Phi+{\widebar Q}^{\dot A}\left({Q}{Q}\right)\Phi\right] .
\end{matrix}
\label{eq:C-4-0}
\end{equation}
We have to prove that this set of terms is equal to the seventh term in the R.H.S. of (\ref{eq:C-1}). This term can be written as
\begin{equation}
\left[\left[\left[{Q}^A,\Phi\right],{Q}_A\right],{\widebar Q}^{\dot A}\right]=\left[\left[{Q}^A,\Phi\right], Q_A\right]{\widebar Q}^{\dot A}-{\widebar Q}^{\dot A}\left[\left[{Q}^A,\Phi\right], Q_A\right],
\label{eq:C-4-1}
\end{equation}
where the $[\,,\,]$ are always intended as graded commutators. The first term on the R.H.S. of (\ref{eq:C-4-1}) can be re-written as 
\begin{equation}
\begin{matrix}
\left(\left[ Q^A,\Phi\right]{Q}_A+{Q}_A\left[{Q}^A,\Phi\right]\right){\widebar Q}^{\dot A}&=&\\&=&\left({Q}^A\Phi{Q}_A-\Phi\,{Q}^A{Q}_A+{Q}_A{Q}^A\Phi-{Q}_A\Phi{Q}^A\right){\widebar Q}^{\dot A}\\
&=&2\left({Q}\Phi{Q}\right){\widebar Q}^{\dot A}-\Phi\left({Q}{Q}\right){\widebar Q}^{\dot A}-\left({Q}{Q}\right)\Phi{\widebar Q}^{\dot A}.
\end{matrix}
\label{eq:C-4-2}
\end{equation}
The second term on the R.H.S. of (\ref{eq:C-4-1}) can be re-written as 
\begin{equation}
\begin{matrix}
-{\widebar Q}^{\dot A}\left(\left[{Q}^A,\Phi\right]{Q}_A+{Q}_A\left[{Q}^A,\Phi\right]\right)&=&\\
&=&-{\widebar Q}^{\dot A}\left({Q}^A\Phi{Q}_A-\Phi{Q}^A{Q}_A\right.\\&+&\left.{Q}_A{Q^A}\Phi-{Q}_A\Phi{Q}^A\right)\quad\quad\\
&=&-{\widebar Q}^{\dot A}\left(\left({Q}\Phi{Q}\right)-\Phi\left({Q}{Q}\right)-\right.\\&-&\left.\left({Q}{Q}\right)\Phi+\left({Q}\Phi{Q}\right)\right)\quad\quad\\
&=&-2\,{\widebar Q}^{\dot A}\left({Q}\Phi{Q}\right)+{\widebar Q}^{\dot A}\Phi\left({Q}{Q}\right)+\\&+&{\widebar Q}^{\dot A}\left({Q}{Q}\right)\Phi.\quad\quad\quad\quad\quad\quad
\end{matrix}
\label{eq:C-5-1}
\end{equation}
Inserting (\ref{eq:C-5-1}) and (\ref{eq:C-4-2}) into (\ref{eq:C-4-1}) we get that it is exactly equal to (\ref{eq:C-4-0}).
\section{}\label{AppendixD}
In this appendix we will give a small sample of the calculations which are needed in order to obtain formula (\ref{eq:29-1}). It would be too long to give all of them. It is easy to prove that the RHS 
of eq. (\ref{eq:29-1}) (with $\alpha$ and $\beta$ not yet specified) can be written as
\begin{equation}
\begin{matrix}
&&e^{-\alpha\,(\theta\theta)\left(\widehat{Q}\widehat {Q}\right)-\beta\,(\bar\theta\bar\theta)\left({\widehat{\widebar Q}}{\widehat{\widebar Q}}\right)}\widetilde \Phi \,e^{\alpha\,(\theta\theta)\left(\widehat Q\widehat {Q}\right)+\beta\,(\bar\theta\bar\theta)\left({\widehat{\widebar Q}}{\widehat {\widebar Q}}\right)}=\quad\quad\quad\quad\quad\quad\quad\quad\quad\\&&
=\widetilde \Phi-\alpha\,(\theta\theta)\left[\widehat{Q}\widehat{Q},\widetilde \Phi\right]
-\beta\,(\bar\theta\bar\theta)\left[\widehat{\widebar Q}\widehat{\widebar Q},\widetilde \Phi\right]+\alpha\beta\,(\theta\theta)(\bar\theta\bar\theta)\left[\widehat{\widebar Q}\widehat{\widebar Q},\left[\widehat{Q}\widehat{Q},\widetilde \Phi\right]\right].\\
\end{matrix}
\label{eq:D-1}
\end{equation} 
Next we should evaluate the various terms on the R.H.S. of (\ref{eq:D-1}). Let us for example calculate the second term $(\theta\theta)\left[\widehat{Q}\widehat{Q},\widetilde \Phi\right]$. 
As it involves a differential operator we should check what the above quantity does when applied to a funcion $F$, i.e.
\begin{equation}
\begin{matrix}
(\theta\theta)\left[\widehat{Q}\widehat{Q},\widetilde \Phi\right]F&=&(\theta\theta)\left(\widehat{Q}^A\widehat{Q}_A\,\Phi-\widetilde \Phi\,\widehat{Q}^A\widehat{Q}_A\right)F\quad\quad\quad\quad\\
&=&(\theta\theta)\left(\widehat{Q}^A\left(\widehat{Q}_A(\widetilde \Phi)F\right)+\widehat{Q}^A\left(\widetilde \Phi\,\widehat Q_A(F)\right)\right.\\
&-&\left.\widetilde \Phi\,\widehat{Q}^A\left(\widehat Q_A(F)\right)\right)\quad\quad\quad\quad\quad\quad\quad\quad\quad\\
&=&(\theta\theta)\left(\widehat{Q}^A\left(\widehat{Q}_A(\widetilde \Phi)\right)F-\widehat Q_A(\widetilde \Phi)\,\widehat{Q}^A(F)+\right.\\
&+&\left.\left(\widehat{Q}^A\widetilde\Phi\right)\widehat{Q}_A(F)+\widetilde \Phi\,\widehat{Q}^A\left(\widehat{Q}_A(F)\right)-\quad\quad\quad\right.\\
&-&\left.\widetilde\Phi\,\widehat{Q}^A\left(\widehat{Q}_A(F)\right)\right)\quad\quad\quad\quad\quad\quad\quad\quad\quad\quad\\
&=&(\theta\theta)\left(\widehat{Q}^A\left(\widehat{Q}_A(\widetilde \Phi)\right)F+2\,\widehat{Q}^A(\widetilde \Phi)\,\widehat{Q}_A(F)\right). 
\end{matrix}
\label{eq:D-2}
\end{equation}
So we get 
\begin{equation}
\begin{matrix}
\left[\widehat{Q}\widehat{Q},\widetilde\Phi\right]=\widehat{Q}\widehat{Q}\,\widetilde \Phi+2\,\widehat{Q}^A(\widetilde \Phi)\,\widehat{Q}_A,
\end{matrix}
\label{eq:D-3}
\end{equation}
and remembering the form of $\widetilde \Phi$ in (\ref{eq:27-1}) we have 
\begin{equation}
\widehat{Q}^A(\widetilde\Phi)=4\,\theta^A\,m-8\,\theta^A\,\bar\theta\bar\theta\,d.
\label{eq:D-4}
\end{equation}
Inserting this in (\ref{eq:D-3}) and multiplying by $(\theta\theta)$ like in the L.H.S. of (\ref{eq:D-2}) we get that the last term on the R.H.S. of (\ref{eq:D-3}) gives zero, so we conclude that :
\[
(\theta\theta)\left[\widehat{Q}\widehat{Q},\widetilde \Phi\right]=(\theta\theta)\,\widehat{Q}\widehat{Q}\,\widetilde\Phi.
\]
Analogously, for the third term in (\ref{eq:D-1}) , we can prove that:
\[
(\bar\theta\bar\theta)\,\left[\widehat{\widebar Q}\widehat{\widebar Q},\widetilde \Phi\right]=(\bar\theta\bar\theta)\widehat{\widebar Q}\widehat{\widebar Q}\widetilde\Phi.
\]
A little more involved is the last term on the R.H.S. of (\ref{eq:D-1}), but the result is: 
\[
(\theta\theta)(\bar\theta\bar\theta)\left[\widehat{\widebar Q}\widehat{\widebar Q},\left[\widehat{Q}\widehat{Q},\widetilde \Phi\right]\right]=(\bar\theta\bar\theta)\,(\theta\theta)\left(\widehat{\widebar Q}\widehat{\widebar Q}\right)\left(\widehat{Q}\widehat{Q}\right)\,\widetilde\Phi.
\label{eq:D-5}
\]
Combining all these terms into the R.H.S. of (\ref{eq:D-1}) and using the expression (\ref{eq:27-1}) of $\widetilde \Phi$, the R.H.S. of (\ref{eq:D-1}) turns out to be: 
\begin{equation}
\begin{matrix}
f(x)-2\,(\theta\theta)\,n(x)\,[1-4\,\alpha]-2\,(\bar\theta\bar\theta)\,m(x)\,[1-4\,\beta]+\\
+4\,(\theta\theta)(\bar\theta\bar\theta)\,d(x)\,[1-4\,\alpha-4\,\beta+16\,\alpha\,\beta].\quad\quad\quad\quad\quad
\end{matrix}
\label{eq:D-6}
\end{equation}
It is easy to see that with the choice $\alpha=\beta=\frac{1}{4}$ the expression (\ref{eq:D-6}) is reduced to just $f(x)$, which is the first term of the original superfield $\Phi$ of (\ref{eq:27-1}).

\section*{Acknowledgments}
The work contained here has been supported by a FRA grant of the University of Trieste and by the  grant GEOSYMQFT  of INFN. This paper is dedicated to the loving memory of Pino Furlan.





\begin{thebibliography}{0}    
\bibitem{ref1.1} J. Wess and B. Zumino, {\it Nucl. Phys. B} {\bf 70}, 39 (1974)
\bibitem{ref1.2} Yu. A. Golfand  and E. P. Likhtman, {\it JETP. Lett. } {\bf 13}, 323 (1971)
\bibitem{ref1.3} D. Volkov and V. Akulov, {\it JETP. Lett. } {\bf 16}, 621 (1972)
\bibitem{ref2.1} J. Wess and B. Zumino, {\it Phys.Lett. B} {\bf 49}, 52 (1974)
\bibitem{ref2.2} J. Iliopoulos and B. Zumino, {\it Nucl. Phys. B} {\bf 76}, 310 (1974) 
\bibitem{ref3} A. Salam and J. Strathdee, {\it Nucl. Phys. B} {\bf 76}, 477 (1974) 
\bibitem{ref4} A. A. Abrikosov Jr., E. Gozzi, and D. Mauro, {\it Ann. Phys.} {\bf 317}, 24 (2005)
\bibitem{ref5.1} E. Gozzi, M. Reuter, and W. D. Thacker, {\it Phys. Rev. D} {\bf 40}, 3363 (1989)
\bibitem{ref5.2} E. Gozzi, {\it Phys.Lett. B} {\bf 201}, 525 (1988)
\bibitem{ref6} R. Abraham and J. Marsden, {\it Foundations of Mechanics}, (Benjamin Press, New York, 1978)
\bibitem{ref7} B. De Witt, {\it Supermanifolds}, (Cambridge University Press, Cambridge, 1987)
\bibitem{ref8} E. Gozzi, M. Reuter {\it Phys. Lett. B} {\bf 233}, 383 (1984)
\bibitem{ref9} H. J. W.  M\"{u}ller-Kirsten and A. Widemann, {\it Supersymmetry: an introduction with conceptual and calculational details}, (World Scientific, Singapore, 1987)

\end{thebibliography}
\end{document}